# Accelerating Drug Repurposing for COVID-19 Treatment by Modeling Mechanisms of Action using Cell Image Features and Machine Learning

Lu Han,[1,+] Guangcun Shan,[2,+,*] Bingfeng Chu,[3] Hongyu Wang,[2,4] Zhongjian Wang,[4] Shengqiao Gao,[1] Wenxia Zhou[1,*]

[1] Beijing Institute of Pharmacology and Toxicology, State Key Laboratory of Toxicology and Medical Countermeasures, Beijing 100850, China;

[2] School of Instrumentation Science and Opto-electronics Engineering & Beijing Advanced Innovation Center for Big Data-based Precision Medicine, Beihang University, Beijing 100083, China;

[3] First Medical Center of PLA General Hospital, Beijing 100853, China

[4] Chengdu Jianshu Technology Co. Ltd, Chengdu 610015, China

*Correspondence should be addressed to

Prof. Dr G.C. Shan. E-mail: gcshan@buaa.edu.cn

Prof. W.X. Zhou, E-mail: zhouwx@bmi.ac.cn

[+]These authors contributed equally to this work.

**Abstract**

The novel coronavirus disease, COVID-19, has rapidly spread worldwide. Developing methods to identify the therapeutic activity of drugs based on phenotypic data can improve the efficiency of drug development. Here, a state-of-the-art machine-learning method was used to identify drug mechanisms of action (MoA) based on the cell image features of 1105 drugs in the LINCS database. As the multi-dimensional features of cell images are affected by non-experimental factors, the characteristics of similar drugs vary considerably, and it is difficult to effectively identify the MoA of drugs as there is substantial noise. By applying the supervised information theoretic metric-learning (ITML) algorithm, a linear transformation made drugs with the same MoA aggregate. By clustering drugs to communities and performing enrichment analysis, we found that transferred image features were more conducive to the recognition of drug MoA. Image features analysis showed that different features play important roles in identifying different drug functions. Drugs that significantly affect cell survival or proliferation, such as cyclin-dependent kinase inhibitors, were more likely to be enriched in communities, whereas other drugs might be decentralized. Chloroquine and clomiphene, which block the entry of virus, were clustered into the same community, indicating that similar MoA could be reflected by the cell image. Overall, the findings of the present study laid the foundation for the discovery of MoAs of new drugs, based on image data. In addition, it provided a new method of drug repurposing for COVID-19.

# 1. Introduction

The emerging coronavirus disease 2019 (COVID-19) has been identified to be caused by the severe acute respiratory syndrome coronavirus 2 (SARS-CoV-2). According to the National Health Commission of The People's Republic of China, since the outbreak of COVID-19 in December 2019 in Wuhan, more than 80 000 patients have been infected in China, and more than 100 million patients have been infected worldwide (WHO, 2019). However, the number of patients diagnosed with COVID-19 and deaths associated with this disease are still increasing. Thus far, there is no proven effective medicine and/or treatment available for COVID-19 (Sanders et al. 2019; Dhawan et al. 2020). Medical teams worldwide have been fully engaged with the COVID-19 pandemic and actively conducting many scientific studies on the pathogenesis, mode of transmission, clinical profiles, management, and disease prevention of this disease (Zhou et al. 2020; Zumla et al. 2020; Li et al. 2019; Wang et al. 2020; Wrapp et al. 2020).

Drug repurposing can help in the rapid identification of potential therapeutic medicines among the existing ones with a known safety profile. Such repurposed drugs might then be promptly used in the clinic to overcome the current therapeutic challenges of COVID-19 (Pushpakom et al. 2019). At present, clinical trials of numerous potential drugs for the treatment of COVID-19 have already begun (Sanders et al. 2019; Dhawan et al. 2020). Apart from drugs that can directly interact with the virus, many drugs may exert antiviral effects through host targets. For example, chloroquine can inhibit the endocytosis of COVID-19 through different mechanisms, such as changing the intracellular environment and increasing the intracellular pH value, to achieve the antiviral effect (Savarino et al. 2003). In addition, some drugs targeting the cell pathway may be effective against SARS-CoV-2 (Kindrachuk et al. 2015).

As virus invasion, replication, and release are highly host dependent, analyzing the effects of drugs on cells is important for the identification of effective antiviral drugs. Transcriptome data, proteome data, and other information obtained from the direct interaction of drugs with intracellular molecules are helpful in the discovery of new therapeutic uses of drugs (Subramanian et al. 2017; Gao et al. 2021). Despite the relatively low cost of obtaining image data, it is difficult to use it for the determination of drug functions owing to the complexity of the data itself and to the fact that it does not directly reflect the molecular characteristics of the drugs. In the present study, the mechanisms of action (MoA) of 1105 drugs on cells were obtained by analyzing their image data. Image data were organized into 812 dimensional vectors, with each dimension representing a specific cell image feature (Nassiri et al. 2018; Corsello et al. 2017), such as Cells_Area, Shape_Area and Cells_AreaShape_Compactness. Owing to the few number of samples and great number of classifications, it is difficult for machine-learning methods to effectively use these data for action-pattern identification. Moreover, because the multi-dimensional features of image data are affected by non-experimental factors, the characteristics of similar drugs vary considerably. In addition, the current sample number is not adequate for a deep-learning model (Aliper et al. 2016; Shan et al. 2020; Gao et al. 2021), and hence, other methods were used for learning optimization. Therefore, we used the supervised information theoretic metric-learning (ITML) algorithm to convert the characteristics of drugs.

By using a non-parametric clustering method (AP cluster), we clustered all drugs into 39 communities and calculated the MoAs of the enriched drugs MoAs in each community. Among the drugs currently being investigated for the treatment of COVID-19, chloroquine and clomiphene could block virus entry by inhibiting endocytosis. Based on the image data of these two drugs, they were both classified to Community 21, despite their different MoA annotations. The analysis of image features showed that drugs from the same community may share similar image features (community-specific image features, CSIFs), which may play an important role in identifying drug functions. In addition, one of the Community 21 members, clomiphene, originally known as an estrogen receptor antagonist, indicated their similar effects on cells. Clomiphene has been reported to block the entry of Ebola virus into the host cell (Nelson et al. 2016).

In the present study, 1105 drugs were analyzed via a machine-learning-based clustering algorithm. A variety of data pre-processing methods were simultaneously used to improve the clustering effect. We compared the principal component analysis (PCA) algorithm and metric-learning algorithm for data pre-processing. For data clustering, we used the affinity propagation (AP) algorithm. Therefore, we adopted the ITML algorithm to perform the identification of drugs with similar mechanisms by optimizing the measurement of drug image features. Compared to the original data and PCA without the ITML algorithm, the new method developed here could help in the identification of drugs with similar mechanisms. For enrichment analysis of drug types with a sample size exceeding five, it was found that 39 clusters were enriched with MoA of 35 drugs. In addition, each community was enriched with drugs with similar MoA: Community 20 was enriched with microtubule inhibitors, tubulin inhibitors, and cyclin-dependent kinase (CDK) inhibitors and Community 21 with drugs known to block virus entry. Further analysis of the image features within each community showed that community-specific image features (CSIFs) may be used to describe and discover the MoA of drugs.

## 2. Methods

*2.1. Data collection and preparing*

The cell morphology data was based on the work of Nassiri and N. McCall (Nassiri and N. McCall 2018).    The cell imaging dataset, containing 1105 drugs of 372 MoAs, was sorted and screened, and 1105 image data, including 812 dimensional image information and encompassing cell responses to the 372 MOAs, were collected (Supplementary Table S1). These 1105 drugs of 372 MoAs covered several clinical uses. The image data represented the most intuitive phenotypic effects of these drugs on cells. The image data included 812 dimensional data, such as Cells_Area, Shape_Area, Cells_AreaShape_Compactness, and Cells_AreaShape_Eccentricity. The distribution of data in each dimension ranged from +677 to −384. We adopted the mean variance normalization method as follows:

Normalezd value = Scale * ((Input - Mean) / sqrt(Variance + Epsilon)) + Bias)

After normalization, the data was distributed within the range of mean = 0, and we collected the information regarding the MoA of the drugs from the LINCS database, which 372 types of MoA. Among these, 49 types were shared by five or more drugs, and the most common MoA was shared by as many as 43 drugs.

*2.2. PCA algorithm*

The PCA algorithm was used to analyze the most important components of input data, and it is often used for data-dimensionality reduction in machine-learning. Through the PCA algorithm, it is possible to reduce an n-dimensional vector to an m-dimensional vector as follows:

$$X(m) = PCA(X(n)), m < n,$$

where X(n) is the original data and X(m) is the output data after mapping the original data from the n-dimensional space to the m-dimensional space. The most representative m-dimensional data were thus extracted from the original data by the PCA algorithm, which can not only reduce the dimensions of input data but also extract the more effective features from the original data.

*2.3. Metric-learning*

In addition to the PCA algorithm, we used a metric-learning method to pre-process input data before clustering. Metric-learning is a machine-learning algorithm for detecting similarities between data and it is widely used in face recognition, for instance. Metric-learning classifies the similarity of input data by learning the distance function in a specific task and it is thus more practical than deep-learning. For example, the deep-learning model trained in a specific task can only adapt to data similar to the training samples, and for input data considerably different from the sample data the results tend to be significantly erroneous. Furthermore, when the number of data categories increases, the former training model needs to be retrained under the new categories, which consumes resources and time. Therefore, practical applications of a deep-learning method are often limited. As a type of machine-learning, metric-learning can effectively solve this problem. It increases the similarity between the same type of data and decreases the similarity between different types of data. Therefore, the result of data clustering after metric-learning is more accurate. In the present study, we used the ITML algorithm of metric-learning to pre-process the input data as follows:

$$S_{n \times n} = ITML(Input_{m \times n}),$$

$$Output_{m \times n} = Input_{m \times n} \times S_{n \times n},$$

where Input$_{m \times n}$ is the input data, m is the number of input data, n is the dimension of input data, S$_{n \times n}$ is the similarity matrix learned by metric-learning, and Output$_{m \times n}$ is the output data after metric learning. In Output$_{m \times n}$, the distance between similar classes of drugs will be closer than that in the original data, and the distance between different classes of drugs will be larger than that in the original data.

*2.4. AP clustering*

After pre-processing, data were clustered. There are different types of clustering algorithms, including the unsupervised and supervised clustering algorithms. The supervised clustering algorithm often needs some prior conditions, such as the categories that need clustering. The unsupervised clustering algorithm often does not require a prior condition, but clustering is performed through the analysis of input data, such as the density or mean of input data. The unsupervised clustering algorithm has a beneficial effect on some data that are difficult to label. In the present study, we used the unsupervised clustering algorithm AP to cluster the data. This algorithm constructed a network for different samples in the input data, and each node in the network represented a sample in the input data. The connection of nodes in this network transferred the responsibility and availability between different samples. After multiple iterations, the AP algorithm generated K exemplars and the remaining samples were allocated to them to complete the clustering. Then, the AP algorithm divided the input data into k categories.

## 3. Results and Discussion

*3.1. Cell imaging dataset containing 1105 drugs*

The cell imaging dataset containing 19,864 unique compounds or drugs were sorted and screened (see Methods section), and image data obtained for 1105 drugs (including 812 dimensional image information), encompassing cell responses to 372 MOAs, were collected (Table S1 in Supplementary Data). These 1105 drugs of 372 MOAs have a broad range of clinical use. The image data represented the most intuitive phenotypic effects of these drugs on cells. The image data comprised 812 dimensional data, including Cells_Area, Shape_Area, Cells_AreaShape_Compactness, and Cells_AreaShape_Eccentricity. The distribution of data in each dimension ranged from +677 to −384, and more than 98.8% of the data were between −20 and 20. We adopted the mean variance normalization method. The data in each dimension followed a normal distribution, with a mean of 0 and a variance of 1, and a range of −7.930 to 13.934. The original data distribution is shown in Figure 1a and the normalized data distribution in Figure 1b.

We collected the MoA information of drugs from the LINCS database, which contained 372 types of MoA for the investigated drugs. Among these, 49 types of MoA were shared by five or more drugs and the most common MoA (adrenergic receptor antagonism) was shared by 43 drugs. The other common MoA were dopamine receptor antagonism, cyclooxygenase inhibition, and serotonin receptor antagonism. The relevant data distribution is shown in Figure 1c for the top ten MoA types; The pie chart represents the MoA. The overall flowchart of the present study is illustrated in Figure 2.

*3.2. Conversion of the 812 dimensional image data by ITML*

The supervised ITML is a global metric-learning method that can be used as an alternative method to understand the metric distance function for a specific task, according to different learning tasks. We used this method to obtain a distance measurement for the drug MoA classification task; parameters were num_constraints (number of constraints to generate) = 20, max_iter (maximum number of iterations) = 1000, and convergence_threshold = 0.001. The t-Distributed Stochastic Neighbor Embedding (t-SNE) plot graphs of the top ten drugs, before and after learning, are shown in Figures 3a and 3b, respectively. Through training for all MoA of drugs, we obtained the T matrix. The 812 dimensional vector was then transformed to a new vector via supervised learning after passing through the T matrix.

*3.3. Classification of drugs into 39 categories based on ITML-transformed features*

We used the T matrix-transformed features to establish drug image phenotype (DIP) connections. The DIP connections were represented as "association scores" computed using Euler distance. To achieve this, for each calculated distance we obtained the corresponding association scores (detailed information is provided in the Methods section and in the Supplementary Distance data file).

The 609,960 pairs of DIP connections (Table S2 in Supplementary Data) observed for the 1105 drugs are shown in the heatmap representation of the distance matrix (Figure S1). The application of an automated, parameter-free clustering algorithm yielded 39 drug groups, with prominent consensus internal DIP similarities. We distinguished each of these 39 groups as a DIP community (Figure 4b). We then used the MoA type composed of more than five drugs as a test set to determine whether the DIP community could be used for drug MoA discovery.

Our enrichment analysis identified significant (P<0.01, Table S3) enriched community-specific drug MoA for each DIP community (Figure 4b and Supplementary Table S4). For example, communities 1, 2, and 3 were enriched with local anesthetics, acetylcholine receptor agonists, and protein kinase A inhibitors, respectively.

To examine whether ITML can help in MoA recognition, we compared the effects of MoA recognition using raw data and data processed by the PCA algorithm and obtained 57 and 48 clusters, respectively. As shown in Table 1, with frequencies of enriched MoA of 26 and 24 and enrichment ratios of 0.4561 and 0.5000, respectively. These were lower than the results of ITML, indicating that clustering of ITML-processed data made it easier to identify drugs with consistent MoA.

### 3.3.1. DIP facilitates identification of drug MoA

Herein, 35 drug MoAs were enriched in the 39 classification communities. Several similar drug MoAs were enriched in the same communities. Protein synthesis inhibitors and histone deacetylase inhibitors were both enriched in Community 18. Cytochrome P450 inhibitors and epidermal growth factor receptor inhibitors were enriched in community 36. Acetylcholine receptor agonists, bacterial cell wall synthesis inhibitors, and angiogenesis inhibitors were enriched in Community 2. Acetylcholine receptor antagonists, retinoid receptor agonists, and tyrosine kinase inhibitors were enriched in Community 4. Adrenergic receptor agonists, norepinephrine reuptake inhibitors, and aromatic hydrocarbon derivatives, for instance, were relatively decentralized and not significantly enriched in all communities. This decentralized distribution may be attributed to the effects of these drugs on the phenotype of tumor cell lines, due to which the image data were not significantly changed. The cell images derived from other cells may be more helpful for the identification of these MoA.

To identify the image features that may be more conducive for the identification of drug use, we calculated the intra-class distance ratio between the features of each dimension in each cluster (Table S5, see the Methods section for details) and determined CSIFs according to the intra-class ratio (<0.01). It was found that the CSIFs rarely overlapped between clusters. Only 26 different features played a role in two clusters, and no features simultaneously became CSIFs in three or more communities. For example, Nuclei_Intensity_MeanIntensity_Ph_golgi was the CSIF of cluster 16 of dopamine uptake inhibitors and of cluster 22, which had no enrichment of any kind of drugs. The CSIFs suggested that drugs within the same cluster may have specific responses to CSIFs. Although there were only four tubulin inhibitors in the dataset, they were all enriched in Community 20, which was also enriched with CDK inhibitors, and only two microtubule inhibitors were in observed in this cluster. The CSIFs corresponding to Community 20 were Cells_Texture_InfoMeas1_Hoechst_5, Cells_Texture_InfoMeas2_Ph_golgi_5, Cells_Texture_Variance_Hoechst_3, and Cytoplasm_AreaShape_Zernike_8_8. This may be related to the effects of the above drugs on the cell cycle, including inhibition of cell division and induction of changes in cell texture. These results suggest that it is feasible to discover the functions of known or new compounds based on DIP (Table S4).

### 3.3.2. Community 21 drugs that could block virus entry

It was found that there are two drugs, chloroquine and clomiphene with different MoA

annotations, were clustered into cluster 21. And these two drug candidates found to be effective against COVID-19. The MoA of clomiphene in cluster 21 was annotated as oestrogen receptor antagonist, which has been found to be resistant to Ebola, suggesting that it may have a similar MoA with chloroqunine. While the MoA of clomiphene was annotated as oestrogen receptor antagonist, different from chloroquine, chloroquine and clomiphene share common drug characteristics. For example, they inhibit T cell proliferation, reduce the release of proinflammatory cytokines, and increase the pH of the endosome to block endocytosis (Savarino et al. 2003; Vincent et al. 2005; Hoffmann et al. 2020). These drugs may be used for COVID-19 prevention and treatment through blocking the PH-dependent pathway. While the SARS-CoV-2 may entry the lung cells via both pH-dependent and pH-independent (TMPRSS2 dependent) pathways and the TMPRSS-2-primed pathway bypassing the endosome-mediated entry may partly explain the low success rates in COVID-19 therapy, chloroquine/hydroxychloroquine alone could not inhibit SARS-CoV-2 infection (Hoffmann et al. 2020). As a result, combination of drugs blocking endocytosis and TMPRSS-2 inhibitors may be promising (Ortega et al. 2020).

The aforementioned results showed that the MoA of drugs can be recognized through multidimensional image features, and the use of ITML for feature conversion may help in the identification of drugs with similar MoA. The DIP communities can help in finding more drugs with similar MoA through the analysis of image data. At present, no effective drug has been approved for COVID-19 treatment. Drug candidates with certain therapeutic effects may be promising, as is the case of chloroquine and remdesivir. Drugs such as remdesivir target viral proteins but the DIP derived from uninfected cell lines may not be able to reflect their MoA.

Chloroquine and clomiphene exert anti-infective effects by regulating the host cell functions. However, serious side effects associated with the use of chloroquine, such as gastrointestinal effects and cardiotoxicity, may limit its clinical use (Doyno et al. 2021). The discovery of new alternative drugs through DIP is of great significance. It was found that the two drugs, chloroquine and clomiphene, in cluster 21 are effective against virus infections. They had different MoA annotations but similar drug characteristics. We observed that adrenergic receptor agonists, norepinephrine reuptake inhibitors, aromatic hydrocarbon derivatives, and drugs with other MoA were not significantly enriched in all the clusters. The effects of these drugs on the phenotype of tumor cells were not significant and may result in image data of non-specific features. Based on the above findings, we suggest that ITML features are more conducive to drug classification than their original features and PCA features.

Image data obtained after the drug acts on the cell is one of the most easily obtained screening data. Evaluating the potential effects of drugs from images is of great significance. Here we used cell characteristic data processed by professional cell image software (CellProfiler) to predict MoA (Carpenter et al. 2006). Due to the complex MoA of drugs, we used third-party MoA annotation data and optimized the metrics based on ITML. The optimized DIP communities were more closely related to the known MoA. Tubulin, CDK, and microtubule inhibitors exert their effects on the formation of spindles, and they were also classified into cluster 20. The effective drug candidates for COVID-19, such as chloroquine, and the anti-Ebola drug clomiphene were accumulated in Community 21. These results confirmed the possibility and accuracy of drug discovery based on image data. It should be noted that the cell image data we used here derived from cell lines without SARS-CoV-2 infection. Therefore, drugs targeting viral proteins may not induce consistent effects on the cells, and hence these data may not be applicable to virus-targeted drug discovery. The image data of SARS-CoV-2-infected cells and under the effects of different drugs would be more useful in screening virus-targeting drugs.

## 4. Conclusions

In summary, we propose a method for discovering MoA based on cell image data after drugs are provided. The functional and association analysis of DIPs provided a hypothesis for drugs' repurposing according to their MoA. The present study indicated that the common mechanism of the drugs under study is related to the DIP observed after they act on the cells. As such, although the other drugs in cluster 21 had different MoA, they may have therapeutic effects similar to those of antiviral drugs within this cluster owing to their similar DIPs. Notably, MoAs, especially at either the target or signaling pathway level, were more suitable as "positive set" labels. However, other drugs contained in cluster 21 had different MoA. The results of the present study showed that the characteristics of ITML conversion were more conducive to the recognition of drug functions. The analysis of feature conversion showed that different features play important roles in identifying different drug functions. With respect to the drugs currently being investigated for COVID-19 treatment, chloroquine and clomiphene showed antiviral effects by inhibiting endocytosis, and were classified into the same community. The MoA of clomiphene in cluster 21 was annotated as estrogen receptor antagonist. As it has also been found to inhibit the entry of Ebola virus (Nelson et al. 2016), and it might have a similar MoA to chloroquine, which was reflected by cell image. Such a combination with drugs block pH-independent pathways may be helpful for COVID-19 treatment. As a matter of fact, the ongoing outbreak of COVID-19 has been overloading medical systems worldwide, and therefore in order to address such a complex challenge, cooperation among diverse researchers with complementary expertise is required (Chen et al. 2020). As the next step, we will be conducting antiviral screening experiments with double-blind clinical trials on some of the predicted drug candidates that could fit the clinical predicting models better. The present work lays the foundation for the discovery of new MoA of drugs based on machine-learning of image data and also provides a new method of drug repurposing for COVID-19 treatment.

**Acknowledgments**

We thank Prof. Baili Zhang and also Dr. Wenke Zheng from Tianjin University of Traditional Chinese Medicine, for their helpful suggestions.

**Author Contributions**

H.L., G.C. S., Z. W. and W.Z. designed the study. H.L., H.Y.W., and G.C. S. performed the experiments and data analyses. H.L., G.C. S., B. Chu, and W.Z. wrote and finalized the manuscript. All authors read and approved the final manuscript.

**Funding** This research work was partially supported by grants from the National Natural Science Foundation of China (No. 81803431) and the Beijing Advanced Innovation Center for Big Data-based Precision Medicine.

**Compliances with ethical standards**

**Conflict of interest**   The authors declare that they have no conflict of interest.

**SUPPLEMENTARY DATA**

The data and codes have been provided in the Supporting file or are available from the authors upon request.

**Table 1: The comparison data for clustering and enrichment.**

|  | Clustering Number | Enrichment Number | Enrichment Expectation (Enrichment Number/ Clustering Number) |
|---|---|---|---|
| Original Data | 57 | 26 | 0.4561 |
| PCA | 48 | 24 | 0.5000 |
| ITML | 39 | 35 | 0.8974 |

**Figure Captions**

Figure 1. Data distribution plots. (a) Original data distribution, (b) Normalized data distribution, and (c) relevant data distribution. The pie chart indicates the MoA. The top 10 MoA types are presented in the bar chart.

Figure 2. Overview of this study approach for drug repurposing analysis. First, we obtained the image data from LINCS (1105 drugs × 812 dimensions). Then, we used PCA and metric-learning (ITML) to process the data. After obtaining the processed data, we used the AP algorithm for clustering data. Finally, we analyzed and compared the results.

Figure 3. t-Distributed Stochastic Neighbor Embedding (t-SNE) plot of (a) original data and (b) ITML-processed data.

Figure 4. Data clustering for (a) original data, (b) ITML-processed data, and (c) PCA data.

**Figure 1**

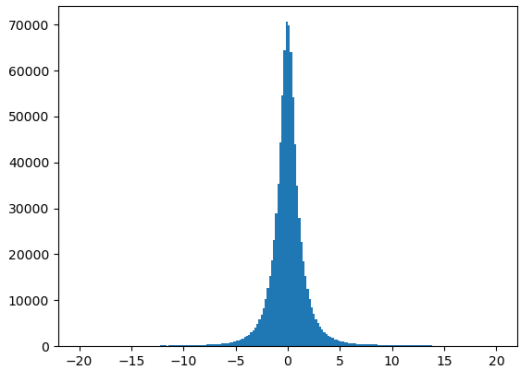
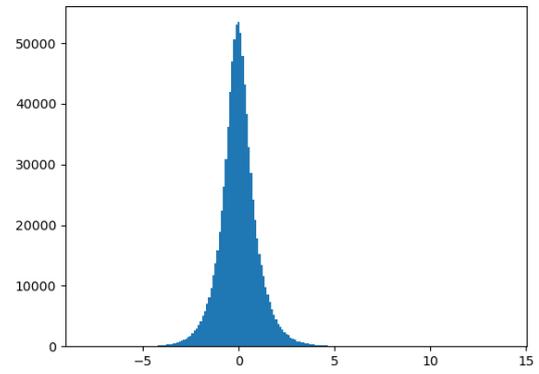

Figure 1a Origin_data   Figure 1b Norm_data

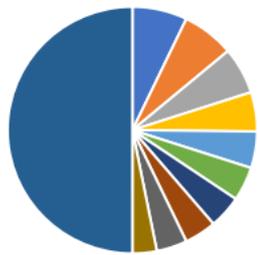
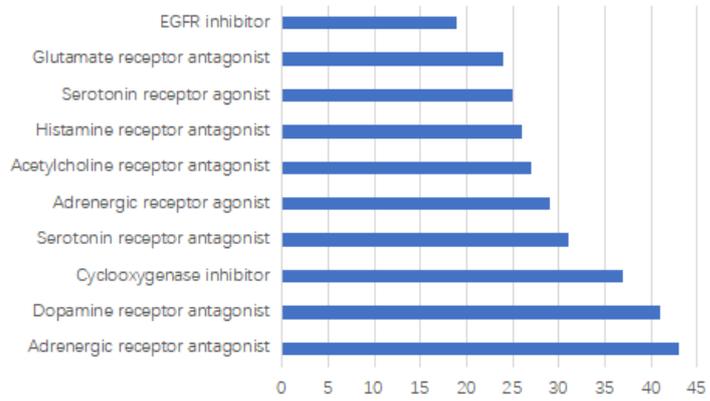

Figure 1c

**Figure 2**

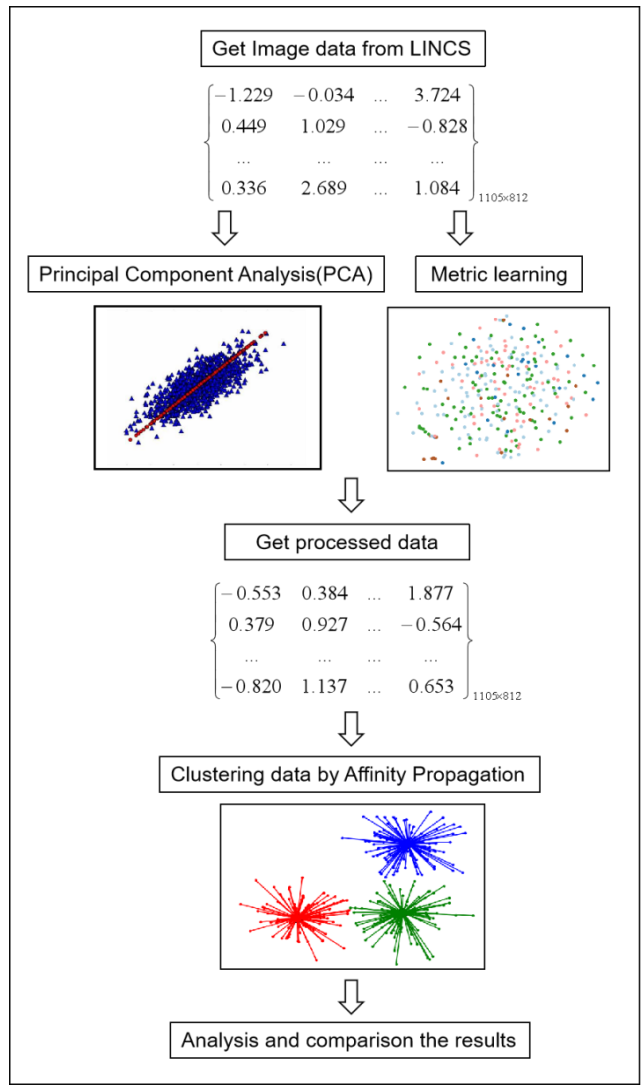

**Figure 3**

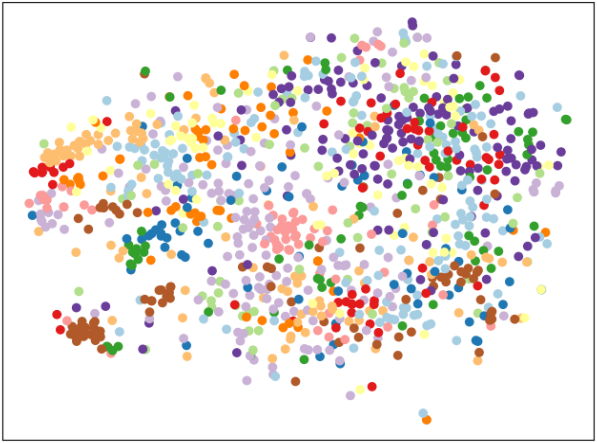 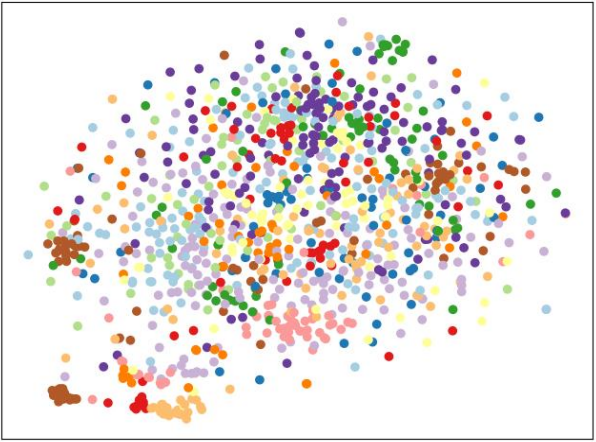

Figure 3a　　　　　　　　　　　　　　　Figure 3b

# Figure 4

**A**

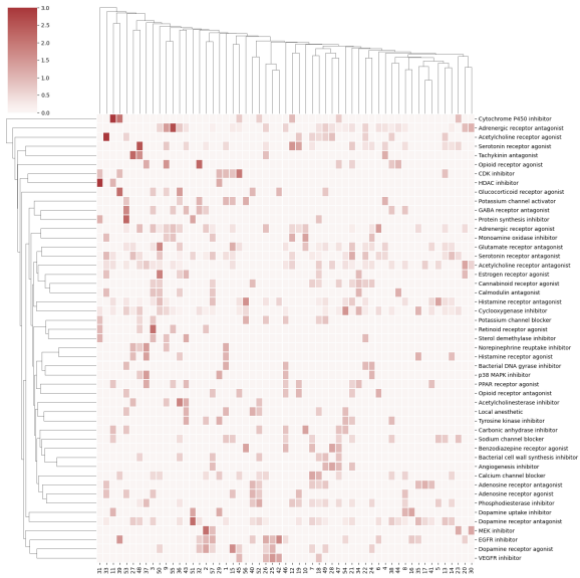

**B**

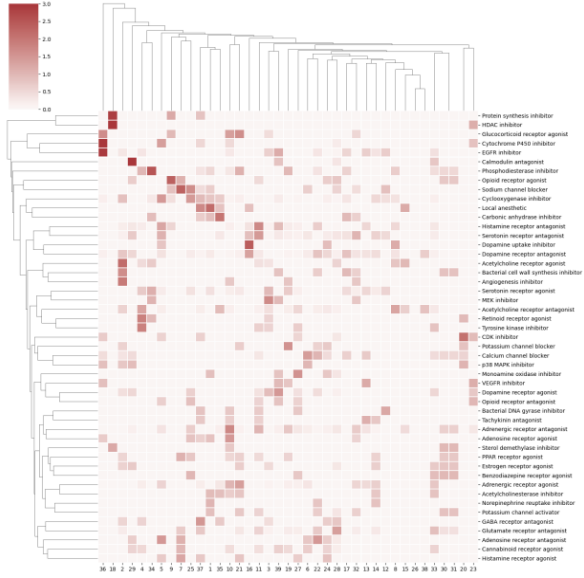

**C**

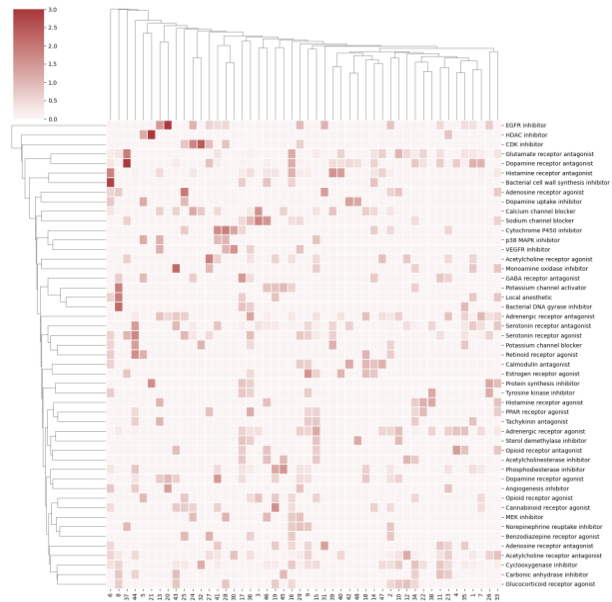